\documentclass[aps,prl,twocolumn,superscriptaddress,showpacs,preprintnumbers,amsmath,amssymb,floatfix]{revtex4}
\usepackage{graphicx}
\usepackage{dcolumn}
\usepackage{bm}
\usepackage[latin1]{inputenc}
\graphicspath{{./figs/}}

\begin{document}

\title{Anisotropic exciton Stark shift in black phosphorus}

\author{A. Chaves} \email{andrey@fisica.ufc.br}
\affiliation{Universidade Federal do Cear\'a, Departamento de
F\'{\i}sica Caixa Postal 6030, 60455-760 Fortaleza, Cear\'a, Brazil}
\author{Tony Low}
\affiliation{Electrical and Computer Engineering Department, University of Minnesota, 200 Union Street SE, 4-174 Keller Hall, Minneapolis, MN 55455-0170}
\author{P. Avouris}
\affiliation{IBM Thomas J. Watson Research Center, Yorktown Heights, NY}
\author{D. \c{C}ak{\i}r}
\affiliation{Department of Physics, University of Antwerp, Groenenborgerlaan 171, B-2020 Antwerp, Belgium}
\author{F. M. Peeters}
\affiliation{Department of Physics, University of Antwerp, Groenenborgerlaan 171, B-2020 Antwerp, Belgium}
\affiliation{Universidade Federal do Cear\'a, Departamento de
F\'{\i}sica Caixa Postal 6030, 60455-760 Fortaleza, Cear\'a, Brazil}

\begin{abstract}
We calculate the excitonic spectrum of few-layer black phosphorus by direct diagonalization of the effective mass Hamiltonian in the presence of an applied in-plane electric field. The strong attractive interaction between electrons and holes in this system allows one to investigate the Stark effect up to very high ionizing fields, including also the excited states. Our results show that the band anisotropy in black phosphorus becomes evident in the direction dependent field induced polarizability of the exciton. 
\end{abstract}

\maketitle

Excitons in semiconductors have been subject of investigation for many years. \cite{Elliot, Kulakovskii} Such interacting electron-hole pair mimics a hydrogen atom, with the hole playing the role of the nucleus. \cite{Koch} It is well known that the hydrogen atom in the presence of an applied electric field exhibits degeneracy breaking and a quadratic energy shift due to the so called Stark effect. The experimental observation of such an effect for excitons in bulk semiconductors is however hindered by the low electron-hole binding energy - for GaAs, for example, this energy is around 4.8 meV, \citep{Pfeiffer} allowing only for very low ionizing electric fields. In such low fields, the exciton binding energy is not significantly modified, albeit the exciton peak broadens due to the decrease of the exciton lifetime. This is a consequence of exciton ionization, which makes the experimental observation of the exciton Stark effect much more challenging. This motivates the study of the Stark effect in quantum wells, which, through the so-called quantum confined Stark effect, can circumvent electron-hole dissociation, allowing the use of higher electric fields. \cite{Miller} Alternatively, the excitonic Stark effect has been also theoretically investigated in carbon nanotubes, \cite{Perebeinos, Avouris} where the electron-hole binding energies, depending on the nanotube configuration, may reach quite large values, \cite{Ando} allowing for high ionizing fields. Nevertheless, setting up an experiment to detect this effect in carbon nanotubes is a difficult task, which has been achieved only very recently. \cite{ExpStark} Strong exciton binding energies are also observed in conjugated polymer chains, \cite{Sebastian} where exciton Stark shifts for high electric fields have been investigated as well. \cite{Weiser2, Weiser}

In recent sudies on single or few-layer semiconductors, such as transition metal dichalcogenides and black phosphorus, exciton binding energies are found to be on the order of hundreds of meV, \cite{Gomez, Chernikov, Hybertsen, Heinz} which brings the possibility of experimentally observing the Stark effect of their excitons. In this context, the case of black phosphorus, \cite{Liu2, Liu, Li2, Koenig, Cakir} a layered material that has recently been fabricated in few-layer form \cite{Gomez} and has a strong potential for technological applications, \cite{Li2, Engel, Buscema, Yuan, Low, Morita} is of special interest. Its effective mass anisotropy \cite{Qiao, Li, Rodin} leads to an exciton wave function with distinct distributions in different in-plane directions, so that the exciton Stark shift behavior, namely, its electric field induced dipole moments and polarizability, is expected to depend on the direction of the applied in-plane electric field, as we will demonstrate by our numerical calculations in what follows. 

We consider the exciton Hamiltonian within the effective mass approximation
\begin{eqnarray} 
H_{exc} = H_e + H_h + V(|\vec r_e - \vec r_h|).
\end{eqnarray}
Due to the high anisotropy of black phosphorus energy bands, it is convenient to write the single particle Hamiltonian in cartesian coordinates $(x_e,y_e,x_h,y_h)$ as
\begin{eqnarray} 
H_i = -\frac{\hbar^2}{2 m_{ix}} \frac{\partial^2}{\partial x_i^2} -\frac{\hbar^2}{2 m_{iy}} \frac{\partial^2}{\partial y_i^2}  + q_{i}\vec{F}\cdot \vec{r_i},
\end{eqnarray}
where $i = e (h)$ represents the electron (hole), $\vec{F}$ is the in-plane applied electric field, $m_{ix(y)}$ is the effective mass of the carrier $i$ in the $x(y)$-direction, and $q_i$ is its charge. It is convenient to rewrite the Hamiltonian in units of the Rydberg energy $R_y$ and Bohr radius $a_0$. We then use relative ($x,y$) and center-of-mass ($\vec R$) coordinates in each direction to simplify the exciton Hamiltonian as
\begin{eqnarray} 
H = -\frac{1}{\mu_{x}} \frac{\partial^2}{\partial x^2} -\frac{1}{\mu_{y}} \frac{\partial^2}{\partial y^2} + V(\sqrt{x^2+y^2}) + e\vec{F}\cdot\vec{r},
\end{eqnarray}
where $\mu_{x(y)}$ is the reduced effective mass in the $x(y)$-direction and the center-of-mass contribution is removed, since the potential does not depend on these coordinates. In order to take proper account of the large dielectric contrast between the vacuum on top of the sample and the substrate below it, the interaction potential $V(r)$, for $r = \sqrt{x^2+y^2}$, is assumed to be of the Keldysh type \cite{Rodin, Keldysh}, which in dimensionless form reads
\begin{equation}\label{eq:pot}
V(r) = -\frac{2\pi}{(\epsilon_1 + \epsilon_2)\rho_0}\left[H_0\left(\frac{r}{\rho_0}\right) - Y_0\left(\frac{r}{\rho_0}\right)\right],
\end{equation}
where $H_0$ and $Y_0$ are Struve and Neumann functions, respectively, $\epsilon_{1(2)}$ is the dielectric constant of the vacuum (substrate) surrounding the phosphorene layer, and $\rho_0 = D\epsilon/(\epsilon_1 + \epsilon_2)$ is the screening factor, where $D$ is the effective width of the layer and $\epsilon$ is its dielectric constant, which is isotropic for our problem of interest, as discussed in Ref. \onlinecite{Low2}. For $n_l$ layers of phosphorene on a SiO$_2$ substrate, this factor was found to be $\rho_0 = n_l10.79$\AA\,. \cite{Gomez} 

To find the excitonic eigenstates, one can use variational functions as proposed in Refs. [\onlinecite{Gomez, Hybertsen}]. However, this kind of approach normally allows one to find only the ground state energy, or just a few excited states. Hence, in order to obtain several exciton states, we rather numerically diagonalize $H$ within a finite difference scheme, which must be performed with a variable mesh, in order to take better account of the singularity of $V(r)$ as $r \rightarrow 0$. Technical details of the finite difference approach in a variable mesh are provided in the supplementary material. \cite{supp}

\begin{figure}[!h]
\centerline{\includegraphics[width=0.9\linewidth]{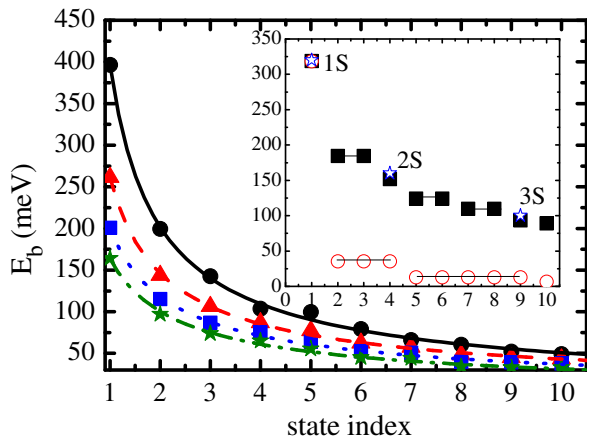}}
\caption{Energies of low-lying exciton states, considering a black phosphorus sample with $n_l$ = 1 (black circles), 2 (red triangles), 3 (blue squares) and 4 (green stars) phosphorene layers. Fitting functions for each case are shown as curves. Inset: exciton energy states for WS$_2$ (black squares) and for an electron-hole pair interacting by the Coulomb potential (red circles) with an effective relative permittivity $\epsilon_r$, along with experimentally obtained values (blue stars, courtesy from the authors of Ref. \cite{Chernikov}), for comparison. Degenerate states are connected by horizontal lines, to help visualization. } \label{fig:excitonenergies}
\end{figure} 

The obtained exciton energies are shown in Fig. \ref{fig:excitonenergies} for four values of thickness (number of phosphorene layers) of the black phosphorus sample, calculated using the same effective masses as in Ref. [\onlinecite{Gomez}]. For the sake of comparison, the inset shows (i) the results for WS$_2$, where conduction and valence bands are known to be isotropic with $\mu = \mu_x = \mu_y = 0.16$; \cite{Hybertsen} (ii) results for the case where the electron and hole interact via the Coulomb potential, i.e. for the 2D hydrogen atom, where the permittivity $\epsilon \approx 5.23 \epsilon_0$ is adjusted as to fit the ground state energy of WS$_2$, by means of the analytical expression $\epsilon^2 = 4R_y\mu/E_0$, where $E_0 =$ 0.318 eV is the ground state exciton binding energy; and (iii) experimental results. \cite{Chernikov}

Let us first discuss the results in the inset. Notice that the energies obtained from the Coulomb potential are clearly underestimated as compared to those calculated with the Keldysh potential. This is a consequence of the fact that the former decays faster than the latter as $r \rightarrow 0$. Moreover, results from the Coulomb potential are analytically found to follow $E_n = - R_y/(n-1/2)^2$, where the states are $(2n+1)$-fold degenerate. In fact, this is easily verified with the numerical method proposed in this work. We point out that binding energies calculated with the actual permittivity of WS$_2$, $\epsilon = 13 \epsilon_0$, would be even more underestimated, since the effective Rydberg energy is inversely proportional to $\epsilon^2$. On the other hand, taking advantage of the fact that the potential in Eq. (\ref{eq:pot}) has a $1/r$ tail for $r \rightarrow \infty$, one can adjust $\epsilon \approx 1.8 \epsilon_0$ as to fit the numerically obtained higher exciton energy states - in this case, due to such smaller effective $\epsilon$, the Coulomb interaction model provides highly \textit{overestimated} binding energies for the lower states. A two-fold degeneracy is still expected due to the circular symmetry of the system, so that positive or negative values of angular momentum $l$ lead to the same energy. However, in the case of the Coulomb potential, additional degeneracies are also observed, due to the SO(3) symmetry of this potential in two dimensions, which is not the case for the Keldysh potential for WS$_2$. Note that for black phosphorus, not even the two-fold degeneracies are observed, since the angular momentum is not a good quantum number in this case, due to the anisotropy of the kinetic energy. 

Inspired by the analytical expression for the eigenstates of the Coulomb potential and regarding the fact that the Keldysh potential in Eq. (\ref{eq:pot}) has the Coulomb form as asymptotic behavior, we fit our numerical results by $E_n = -\gamma_{1}/(n + \gamma_2)^{\gamma_3}$, as shown by the curves in Fig. \ref{fig:excitonenergies}. Fitting parameters are given in Table I. The ground state exciton energies for $n_l = $1, 2, 3 and 4 phosphorene layers are found to be -396, -261, -200 and -163 meV, respectively. This is in good agreement with (namely, less than 10$\%$ lower than) the results obtained by the variational approach. \cite{Gomez} The three lowest energy $s$ states of WS$_2$, are found to be -318.80, -151.72, and -93.56 meV, also in good agreement with recent experimental results for this material, which are shown as open blue stars in the inset of Fig. \ref{fig:excitonenergies}, \cite{Chernikov} which validates our method. 

\begin{table}
\caption{Fitting parameters for the exciton spectra in Fig. \ref{fig:excitonenergies}, using the Rydberg-like expression $E_n = \gamma_{1}/(n + \gamma_2)^{\gamma_3}$, considering different number of phosphorene layers. The parameter $\gamma_1$ is in units of meV, whereas $\gamma_2$ and $\gamma_3$ are dimensionless.} \label{tab:2}\renewcommand{\arraystretch}{1.5}
\begin{ruledtabular}
\begin{tabular}{ccccc}
 $n_l$ & 1 & 2 & 3  & 4   \\\hline
 $\gamma_1$ & 317.38 & 223.92 & 183.51 & 170.45 \\
 $\gamma_2$ & -0.24 & -0.19 & -0.119 & -0.0586 \\
 $\gamma_3$ & -0.80 & -0.72 & -0.699 & -0.733 \\
\end{tabular}
\end{ruledtabular}
\end{table}

By using Fermi's golden rule, one can calculate the probability for a photon induced valence band-to-exciton transition to occur, which is demonstrated to be proportional to $|\langle c|\hat{e}\cdot\vec{p}|v\rangle|^2|\psi_n(0,0)|^2$, where the first factor is the dipole matrix element between conduction and valence band states. The optically active transition in the case of light polarized in the $x-$direction is between the band states labeled as $\Gamma_2^+$ and $\Gamma_4^-$ in Ref. \onlinecite{Li}, which is the one we deal with here. In fact, such a polarization in the lighter effective mass direction has been recently verified experimentally. \cite{XWang} With this factor set, we just need now to investigate the squared modulus of the exciton envelope function at $r = 0$, i.e. $|\psi_n(0,0)|^2$, as being the oscillator strength. Indeed, it is reasonable that electron hole recombination processes are more likely when electron and hole are in the same place in real space, i.e., if $|\psi_n(0,0)|^2$ is non-zero. Therefore, for isotropic materials such as WS$_2$, the optically active exciton states are only those with $s$-like wavefunctions. This means that the exciton Rydberg series observed in the experiments performed in Ref. [\onlinecite{Chernikov}] are not the complete exciton energy series: $l \ne 0$ states are missing in the spectrum, and only the states labelled as $1s$, $2s$ and $3s$ in the inset of Fig. \ref{fig:excitonenergies} are observed experimentally in e.g. Fig. 3 of Ref. [\onlinecite{Chernikov}], with very good quantitative agreement. The $p$ states would appear in two-photon experiments, yet with a different energy series, in contrast with the case of Coulomb-like electron-hole interaction, where for each $p$ state there is a degenerate $s$ state. In fact, the $p$ states in between 1$s$ and 2$s$ states observed in our results are consistent with recent experimental observations. \cite{Ye} In the case of black phosphorus, the angular momentum is no longer a good quantum number, since the kinetic energy operator is not circularly symmetric, but we observe that $n = 1$ ($E_1$ = -396.37 meV) and $n = 3$ ($E_3$ = -142.78 meV) states in the spectrum of Fig. \ref{fig:excitonenergies} are optically active, i.e. $|\psi_1(0,0)|^2 \neq 0$ and $|\psi_3(0,0)|^2 \neq 0$, due to $s$-like orbital nature of their wave functions (details of the wave functions for these and other exciton states are shown in the supplementary material). Thus, we will refer only to these states in the discussion about the exciton Stark effect from here onwards.

\begin{figure}[!h]
\centerline{\includegraphics[width=\linewidth]{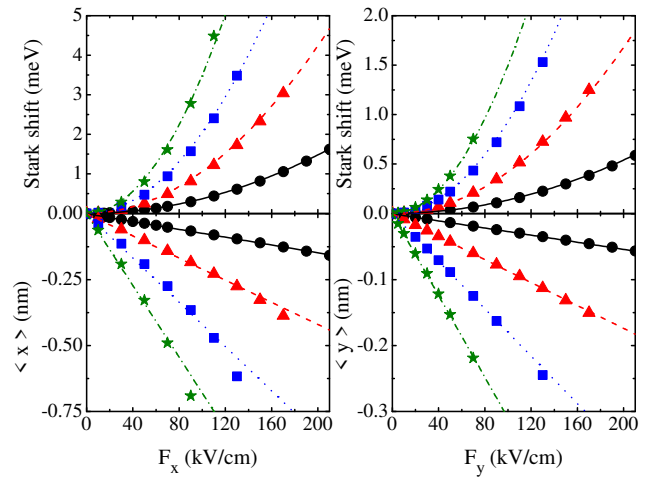}}
\caption{Ground ($n$ = 1) state exciton stark shift, considering electric fields applied in $x$ (left panels) and $y$-direction (right panels), for systems with $n_l = $ 1 (black circles), 2 (red triangles), 3 (blue squares) and 4 (green stars) phosphorene layers. Expectation values of the electron-hole polarization as a function of the applied electric field intensity are shown in the bottom panels. Curves in top and bottom panels are quadratic and linear fittings for the stark shift and polarization, respectively.} \label{fig:Stark}
\end{figure} 

Due to the anisotropic band structure of black phosphorus, some physical properties of this material are expected to be direction dependent. Exciton response to applied electric fields, for instance, should be stronger in the direction where the reduced effective mass is lower. This is indeed observed in Fig. \ref{fig:Stark}, which shows, as symbols, the numerically obtained Stark shift for an electric field applied in $x$ (left panels) and $y$-directions (right panels), considering $n_l = 1$ - 4 layers. The quadratic Stark shift follows the expression 
\begin{equation}\label{eq.stark}
\Delta E = E(F) - E(F = 0) = pF + \beta F^2
\end{equation}
where $F$ is the magnitude of the electric field, $p$ is the intrinsic dipole moment, and $\beta$ is the polarizability parameter. Such a quadratic behavior is, to a good extent, observed in our numerical results, which are well fitted by the quadratic curves in the top panels of Fig. \ref{fig:Stark}, where we assumed $p = 0$ and used $\beta$ as a fitting parameter in Eq. (\ref{eq.stark}). The $p = 0$ assumption is justified by the fact that we are dealing only with non-degenerate states, which do not possess a permanent dipole moment, so that the linear Stark shift can be neglected. The quadratic Stark shift comes from the fact that the applied field is able to induce electron-hole polarization, which depends linearly on the strength of the applied field. Such linear behavior is indeed observed in the bottom panels of Fig. \ref{fig:Stark}, where the numerically obtained polarizations (symbols), i.e. the expectation values of the electron-hole separation $\langle x \rangle$ and $\langle y \rangle$, are well fitted by linear functions (curves). As observed in Fig. \ref{fig:excitonenergies}, the ground state exciton energy decreases as the number of layers increases. As a consequence, the exciton energy becomes more succeptible to electric field effects for higher number of layers and the Stark shift is more clear in these cases - for instance, considering $n_l = 4$, the exciton ground state exhibits a Stark shift of $\approx$ 4.5 meV and a polarization of $\langle x \rangle \approx 25$ \AA\, for fields as high as $F = 100$ kV/cm applied in the $x$ direction, whereas a much smaller $\Delta E \approx$ 1.5 meV and $\langle x \rangle \approx 4.5$ \AA\, is observed for $n_l = 1$ in a field twice higher. Nevertheless, even such a small Stark shift is still within the range of energies that are detectable with current experimental techniques (see, e.g. Ref. \onlinecite{Gerardot}). It is also clear that electric fields applied in $x$-direction affect more the exciton polarization and, consequently, lead to higher Stark shift as compared to a field applied in $y$-direction.

Another interesting feature of the exciton Stark effect in black phosphorus, as compared to the one in conventional bulk semiconductors, is the opportunity to observe such effect also in an excited state. The same study we made for the ground state exciton in Fig. \ref{fig:Stark} was also done for the second excited state ($n$ = 3), where quite similar features are observed, although small deviations from quadratic Stark shift are observed at high applied fields, which also occurs for carbon nanotubes exciton Stark shift. \cite{Perebeinos} A clear difference, however, lies in the stronger Stark shift observed in the $n = 3$ case, as compared to the previous case, which is due to its lower binding energy: in the $n_l = 4$ case, for instance, a $ \Delta E \approx$17 meV shift is obtained for an electric field of 50 kV/cm applied in the $x$-direction. A similar shift is also obtained for $n_l = 1$, but for a higher field, $F_x = 110$ kV/cm. On the other hand, such lower binding energy also forbids us to investigate the effect up to higher fields without dissociating the electron-hole pair. The polarizabilities of the exciton ground and second excited states, for systems with different number of layers, are summarized in Table \ref{tab:1}, where the anisotropy of the Stark effect becomes quite evident by the distinct polarizability parameters found for electric fields applied in different directions.
\begin{table}
\caption{Polarizability parameters $\beta^{n}_{x}$ and $\beta^{n}_{y}$ for the $n = 1$ and 3 states, in units of e\AA$^2$/mV, for electric fields applied in the $x$ and $y$ direction, respectively.} \label{tab:1}\renewcommand{\arraystretch}{1.5}
\begin{ruledtabular}
\begin{tabular}{ccccc}
 $n_l$ & 1 & 2 & 3  & 4   \\\hline
 $\beta^1_{x(y)}$ & 0.37 (0.135) & 1.06 (0.42) & 2.05 (0.93) & 3.5 (1.5) \\
 $\beta^3_{x(y)}$ & 10 (4.8) & 27 (13) & 48 (23) & 65 (43) \\
\end{tabular}
\end{ruledtabular}
\end{table}

\begin{figure}[!h]
\centerline{\includegraphics[width=\linewidth]{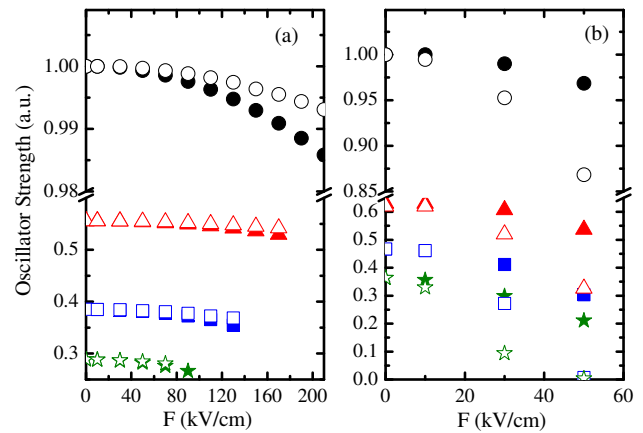}}
\caption{Oscillator strengths for the (a) ground ($n = 1$) and (b) second excited ($n = 3$) exciton states as a function of the strength $F$ of an electric field applied in the $x$ (closed symbols) and $y$ (open symbols) directions. The symbols refer to the different number of layers, as in Fig. 1.} \label{fig:StarkOS}
\end{figure}

The polarization induced by the field is expected to contribute to a broadening of the excitonic peak in the absorption spectrum and, therefore, hinder the visualization of the Stark effect. Nevertheless, our results for the oscillator strength in Fig. \ref{fig:StarkOS} suggest that such contribution is quite small in the ground state, especially for a monolayer: $|\psi_1(0,0)|^2$ changes by only $\approx 2 \%$ (1$\%$) for fields up to $F_{x(y)} = 210$ kV/cm, whereas such a change is already observed for a field of $F_x = 90$ ($F_y = 70$) in the case of four layers. As for the excited state, $|\psi_3(0,0)|^2$ changes up to $\approx 14 (27)\%$ in the case of $F_y = 50$ (30) kV/cm for $n_l = 1 (4)$, as compared to the zero field case. Results in each panel are divided by the zero field oscillator strength of the $n_l = 1$ case. This means that the $n_l = 2, 3$ and 4 systems exhibit ground (second excited) state oscillator strengths that are  $\approx$ 55.5$\%$ (61.9$\%$), 38.5$\%$ (46.7$\%$), and 28.8$\%$ (36.3$\%$) of that of the monolayer case, respectively. Notice that the influence of the electric field on the ground state oscillator strength is higher when the field is applied in the $x$-direction (closed symbols), whereas for the $n = 3$ state, it is higher for $F$ applied in the $y$-direction (open symbols). This can also be understood as a manifestation of the anisotropy of the exciton wave function: although $n = 1$ and 3 are both $s$-like states, we observe that $\langle x^2 \rangle > \langle y^2 \rangle$ for the former, whereas $\langle x^2 \rangle < \langle y^2 \rangle$ for the latter (cf. supplementary material), which explains such opposite behaviour of the oscillator strengths of these states under applied fields.

In conclusion, we have calculated the excitonic states in few layer black phosphorus in the presence of an external electric field, by numerical diagonalization of the effective mass Hamiltonian. The method developed here is shown to be easily adapted for other layered materials and agrees well with recent experiments on WS$_2$. Due to the non-Coulomb effective interaction potential between electron and hole in such 2D system, the degeneracy coming from the SO(3) symmetry of the planar hydrogen atom model is lifted, whereas the bands anisotropy is responsible for lifting the degeneracy of angular momentum eigenstates. Even so, the exciton spectrum in black phosphorus can still be satisfactorily fitted by a hydrogen-like expression. Due to the high binding energies, excitons can withstand strong in-plane electric fields without dissociating. \cite{footnote} This allowed us to observe Stark shifts up to $\approx$15 meV, with electric fields up to $\approx$200 kV/cm, not only for the ground state exciton, but also for excited states, without significant depreciation of the oscillator strength. We believe that the clear and unusually anisotropic exciton Stark shift predicted here may stimulate further photoluminescence and reflectance experiments in few layer black phosphorus under in-plane electric fields in the near future.

\acknowledgements Discussions with J. M. Pereira Jr. and J. S. de Souza are gratefully acknowledged. This work was supported by the Brazilian Council for Research (CNPq), the Flemish Science Foundation (FWO-Vl), the Methusalem programme of the Flemish government, and the Bilateral program (CNPq-FWO) between Flanders and Brazil. D.C. is supported by a FWO Pegasus-short Marie Curie Fellowship.


\begin{thebibliography}{apsrev}
\bibitem{footnote} Since the electric field potential diverges to $-\infty$ as $r \rightarrow \infty$, any electron-hole pair is expected to eventually dissociate. However, such dissociation is highly unlikely to occur before the exciton decays, since it would require tunneling through a significant thick potential barrier, provided by the strong electron-hole interaction potential.
\end{thebibliography}
\end{document}


\title{Anisotropic exciton Stark shift in black phosphorus: Supplementary material}
\author{A. Chaves} \email{andrey@fisica.ufc.br}
\affiliation{Universidade Federal do Cear\'a, Departamento de
F\'{\i}sica Caixa Postal 6030, 60455-760 Fortaleza, Cear\'a, Brazil}
\author{Tony Low}
\affiliation{Electrical and Computer Engineering Department, University of Minnesota, 200 Union Street SE, 4-174 Keller Hall, Minneapolis, MN 55455-0170}
\author{P. Avouris}
\affiliation{IBM Thomas J. Watson Research Center, Yorktown Heights, NY}
\author{D. \c{C}ak{\i}r}
\affiliation{Department of Physics, University of Antwerp, Groenenborgerlaan 171, B-2020 Antwerp, Belgium}
\author{F. M. Peeters}
\affiliation{Department of Physics, University of Antwerp, Groenenborgerlaan 171, B-2020 Antwerp, Belgium}
\affiliation{Universidade Federal do Cear\'a, Departamento de
F\'{\i}sica Caixa Postal 6030, 60455-760 Fortaleza, Cear\'a, Brazil}

\maketitle

In order to find the eigenstates of the Hamiltonian Eq. (3) of the main manuscript, we propose a finite difference scheme on a variable mesh, which allows us to take better account of the singularity of the electron-hole interaction potential as the relative coordinate $r \rightarrow 0$. The $xy$-plane is discretized in $N_x\times N_y$ points (here, we take $N_x = N_y = 1020$) separated by $h^{x(y)}_j = x(y)_{j}-x(y)_{j-1}$, where the mesh varies according to $h^{x(y)}_{j} = \Delta^{x(y)}_{min} + \Delta^{x(y)}_{max}\left\{1 - \exp\left[-(j-N_{x(y)}/2-1)/\delta^{x(y)}\right] \right\}$. This sets the values of $h^{x(y)}_j$ for $j > N_{x(y)}/2$, namely, for $x(y) > 0$. The grid is then mirror reflected to the negative part of the $x(y)$-axis. In this way, we can control how close the points are to $r = 0$ by adjusting $\Delta^{x(y)}_{min}$, tune how narrow the mesh is in the neighborhood of this point by adjusting $\delta^{x(y)}$, and set a larger mesh far from $r = 0$ by adjusting $\Delta^{x(y)}_{max}$. Results in this paper are obtained with  $\Delta^{x(y)}_{min} = 5 \times 10^{-5}$ \AA\,, $\delta^{x(y)} = 100$ and $\Delta^{x(y)}_{max} = 0.4$\AA\,.

After discretization, we write the two-dimensional Hamiltonian as a five-diagonals matrix and numerically diagonalize it. The grid points, labeled in the (line, column) form $(i,j)$, are then mapped into a single label $l$. For a non-uniform mesh, diagonalizing this matrix is cumbersome, because the discretized Hamiltonian matrix is asymmetric:
\begin{eqnarray}
H_{l,l} = \frac{1}{h^{x}_{j+1}L^2_{x,j}} + \frac{1}{h^{x}_{j}L^2_{x,j}} + \frac{1}{h^{y}_{i+1}L^2_{y,i}} + \frac{1}{h^{y}_{i}L^2_{y,i}} + V_l,   \nonumber \\
H_{l,l+1} = -\frac{1}{\mu_y h^{y}_{i+1}L^2_{y,i}}, \quad H_{l,l-1} = -\frac{1}{\mu_y h^{y}_{i}L^2_{y,i}}, \quad\quad  \nonumber \\
H_{l,l+N_y} = -\frac{1}{\mu_x h^{x}_{j+1}L^2_{x,j}},  
\quad H_{l,l-N_y} = -\frac{1}{\mu_x h^{x}_{j-1}L^2_{x,j}}, \quad
\end{eqnarray}
where $L^2_{x(y),j(i)} = \left(h^{x(y)}_{j(i)+1}+h^{x(y)}_{j(i)}\right)/2 $. Nevertheless, symmetrization of the Hamiltonian can be achieved by multiplying all the terms in the Hamiltonian by $L^2_{x,j}L^2_{y,i}$, or equivalently, $\mathcal{B} = \mathcal{L}\mathcal{L}H$,
where $\mathcal{L}$ is a diagonal matrix whose diagonal elements are $L_{x,j}L_{y,i}$. Since $\mathcal{L}\mathcal{L}H \mathcal{L}^{-1}\mathcal{L}\psi = E \mathcal{L}\mathcal{L}\psi$, then $\mathcal{B}\mathcal{L}^{-1}\mathcal{L}\psi = E \mathcal{L}\mathcal{L}\psi$ or, equivalently $\mathcal{L}^{-1}\mathcal{B}\mathcal{L}^{-1}\Phi = E \Phi$ (where $\Phi = \mathcal{L}\psi$). Hence, diagonalizing $\mathcal{L}^{-1}\mathcal{B}\mathcal{L}^{-1}$, which is symmetric, gives the same eigenenergies as diagonalizing the Hamiltonian itself. This matrix is then numerically diagonalized by Arnoldi's method in order to obtain the exciton states. 

\begin{figure}[!h]
\centerline{\includegraphics[width=\linewidth]{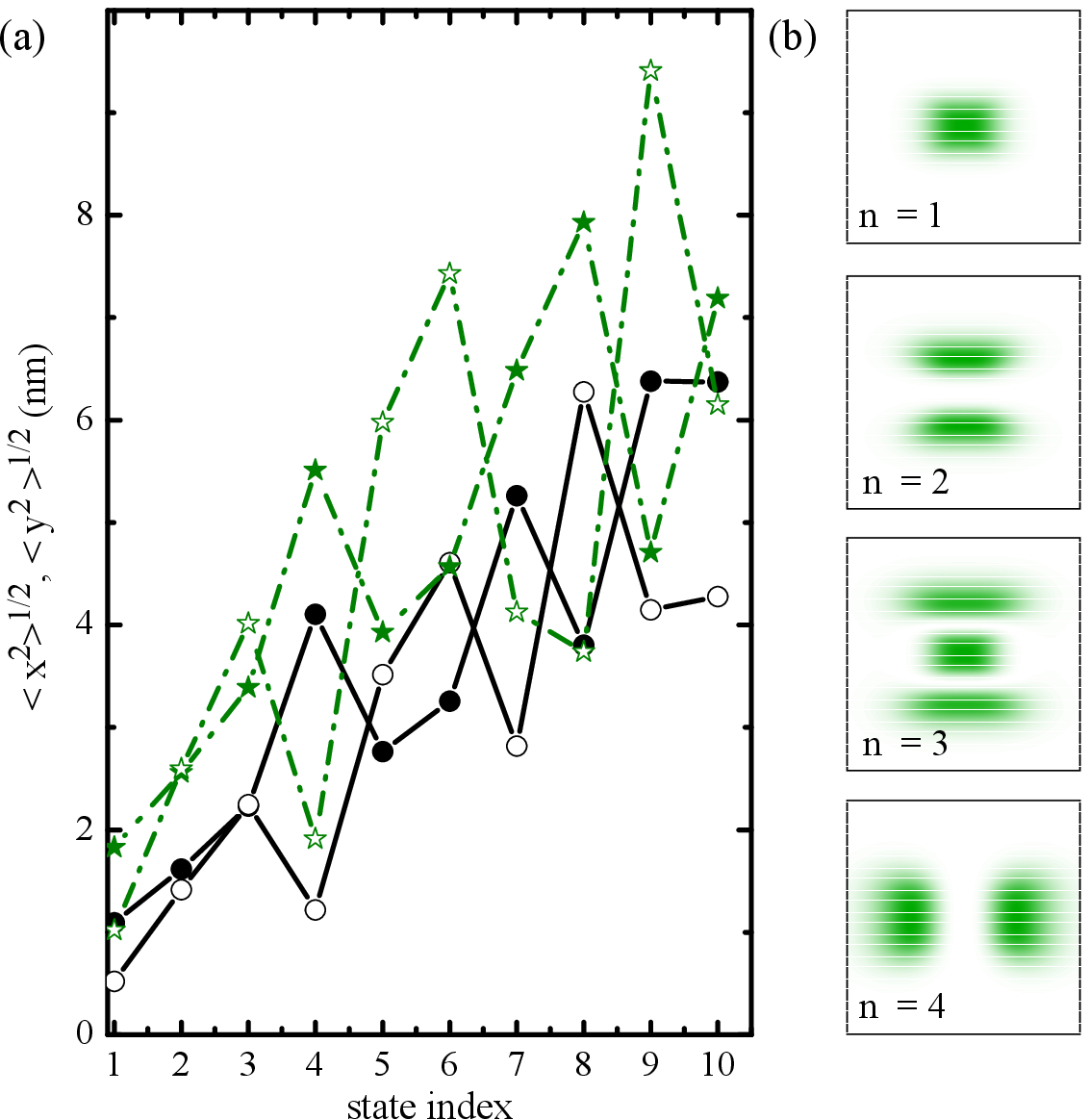}}
\caption{(a) Average width of the exciton function in each direction for $n_l$ = 1 (black circles) and 4 (green stars). Open (full) symbols are results for $\langle y^2 \rangle$ ($\langle x^2 \rangle$). (b) Wave functions for the four low-lying exciton states, considering $n_l = 4$ black phosphorus in the absence of electric fields.} \label{fig:excitonenergies}
\end{figure} 

The exciton energies obtained from this method are shown in Fig. 1 of the main manuscript. As for the eigenfunctions, Fig. \ref{fig:excitonenergies}(a) shows the width of each exciton state wave function, considering $n_l$ = 1 (black circles) and 4 (green stars). Linear interpolating curves are shown just as a guide to the eye. The anisotropy (namely, the difference between $\langle x^2 \rangle$ (open symbols) and $\langle y^2 \rangle$ (full symbols)) is more evident for some specific exciton states. As verified by the wave functions for the four low-lying exciton energy states in $n_l = 4$ case, shown in Fig. \ref{fig:excitonenergies}(b), the $n = 1$ state resembles a $1s$ orbital, as expected, but its width is twice larger in the direction of lighter effective mass, namely, the $x$-direction. Subsequent states resemble $2p_y$, $2s$ and $2p_x$ orbitals, respectively. Similar wave functions are also found in Fig. 7 of Ref. \cite{Rodin} for $n_l = 1$. The $n = 2$ state exhibits $\langle x^2 \rangle \sim \langle y^2 \rangle$, which is unusual for a $p_y$ state, that is normally more spread in the $y$-direction. However, this is compensated in the case of phosphorene by its lighter effective mass in the $x$-direction. The $n = 3$ state, which is a $2s$-like orbital, exhibits a slightly higher width in the $y$-direction, which is more evident in the $n_l = 4$ case, as compared to $n_l = 1$. This has consequences for its oscillator strength, as discussed in the main manuscript. From the widths shown in Fig. \ref{fig:excitonenergies}(a), one can infer that the exciton wave function in both directions is spread over tens of angstrons, i.e. over a region many times larger than the inter-atomic distance $a \approx 2 $\AA\, in black phosphorus, \cite{Gomez} as required for the validity of the Wannier-Mott theory of excitons used here. In fact, just like in carbon nanotubes, \cite{Nguyen, AvourisCNT} excitons in few layer black phosphorus undergo low screening outside the layer, resulting in quite large electron-hole binding energies, as opposed to those of Frenkel excitons. Nevertheless, the dielectric function in the layer itself is on the same order of magnitude of those of other semiconductor materials, \cite{Gomez} such as Si and GaAs (namely, $\epsilon \approx 10 \epsilon_0$), which leads to the large spatial extension of the exciton wave function observed here, thus justifying the use of Wannier-Mott theory.